# Network Traffic Forensics on Firefox Mobile OS: Facebook, Twitter and Telegram as Case Studies

**Mohd Najwadi Yusoff**
School of Computer Science, Universiti Sains Malaysia, Penang, Malaysia.
mohd.najwadi@gmail.com

**Ali Dehghantanha**
School of Computing, Science and Engineering, University of Salford, Manchester, United Kingdom.
A.Dehghantanha@Salford.ac.uk

**Ramlan Mahmod**
Faculty of Computer Science and Information Technology, Universiti Putra Malaysia, Serdang, Malaysia.
ramlan@upm.edu.my

## Abstract

Development of mobile web-centric OS such as Firefox OS has created new challenges, and opportunities for digital investigators. Network traffic forensic plays an important role in cybercrime investigation to detect subject(s) and object(s) of the crime. In this chapter, we detect and analyze residual network traffic artefacts of Firefox OS in relation to two popular social networking applications (Facebook and Twitter) and one instant messaging application (Telegram). We utilized a Firefox OS simulator to generate relevant traffic while all communication data were captured using network monitoring tools. Captured network packets were examined and remnants with forensic value were reported. This paper as the first focused study on mobile Firefox OS network traffic analysis should pave the way for the future research in this direction.

# 1. Introduction

The significant rise of social media networking, instant messaging platform, webmail and other mobile web applications has spawned the idea to build mobile web-centric operating systems (OS) using different open web standards like HTML5, CSS3, and JavaScript. Due to that fact, Mozilla has released the world's first mobile web-centric OS on February 21st, 2013. This mobile web-centric OS is based on Firefox web browser rendering engine on top of Linux kernel, called Firefox OS (FxOS) (Mozilla Corporation, 2013). The emergence of mobile web-centric OS has created new challenges, concentrations and opportunities for digital investigators. In general, the growth of mobile devices may allow cybercriminals to utilize social media and instant messaging services for malicious purposes (Mohtasebi and Dehghantanha, 2011) such as spreading malicious codes, obtaining and disseminating confidential information etc. Furthermore, copyright infringement, cyber stalking, cyber bullying, slander spreading and sexual harassment are becoming serious threats to mobile device users (Dezfouli et al., 2013). Therefore, it is common to confront with different types of mobile devices during variety of forensics investigation cases (Damshenas et al., 2014).

Many previous studies were focused on detection and analysis of network traffic artefacts for cyber forensics. Quick and Choo run Dropbox cloud storage in virtual environment machine and all network activities were recorded (Quick and Choo, 2013a). Network Miner 1.0 (Hjelmvik, 2014) and Wireshark Portable 1.6.5 (Combs, 2013) were used to capture the network traffic in many circumstances and network traffic was seen on TCP port 80 and 443 only. Quick and Choo also run Microsoft SkyDrive in virtual environment machine using the same method (Quick and Choo, 2013b). The result were tabled with more information such as IP start, IP finish, URL observed in network traffic and registered owner. No username and password were observed in the clear text network traffic. Quick and Choo continued the research and run Google Drive cloud storage as the case study (Quick and Choo, 2014). Google Drive account credential was observed but cannot be seen in the network traffic, suggesting the data was encrypted. Martini and Choo observed network traffic from virtual environment network adapter using ownCloud as a case study (Martini and Choo, 2013). HTTP Basic authentications were captured and user's ownCloud credentials were successfully displayed. Farina produced an analysis of the sequence of network traffic and file I/O interactions in the torrent synchronization process (Farina et al., 2014). Bittorrent client were used as the test subject.

Utilizing virtual machines for generating network traffic is a very common in forensics research. Blakeley investigate cloud storage software using hubiC as a case study (Blakeley et al., 2015). In the network analysis part, it was observed that a redirect to HTTPS on port 8080 was returned when the initial request was made to www.hubic.com. This shows that there was no plaintext traffic accepted by the hubiC website. Shariati run network analysis using SugarSync cloud storage (Shariati et al., 2015). Majority of communication were encrypted and no credentials or contents of sample data-set were able to be recovered. Dezfouli investigated Facebook, Twitter, LinkedIn and Google+ artefacts on Android and iOS. During experiment, network activities were captured and able to determine user's IP address, domain name of connected social media sites and corresponding session timestamps. Yang identify network artefacts of Facebook and Skype Windows Store application (Yang et al., 2016). Yang was able to correlate the IP addresses with the timestamp information to determine when the application was started up and the duration of application used during experiment. Daryabar investigated OneDrive, Box, Google Drive, Dropbox applications on Android and iOS devices (Daryabar et al., 2016a). In network analysis part, the connection were secured between cloud client application and the server, thus no credential were able to be seen. Daryabar also investigated the MEGA cloud client application on Android and iOS (Daryabar et al., 2016b). Daryabar identify network artefact arising from user activities, such as login, uploading, downloading, deletion, and file sharing including timestamps. Table 1 is reflecting literature summary of network analysis and monitoring captured through virtual environment network adapter.

Table 1 - Summary of Network Analysis and Monitoring Captured through Virtual Environment Network Adapter

| Researcher(s) | Application(s) | Application Type |
|---|---|---|
| **Martini and Choo 2013** | OwnCloud | Cloud Storage |
| **Quick and Choo 2013b** | Dropbox | Cloud Storage |
| **Quick and Choo 2013c** | Microsoft SkyDrive | Cloud Storage |
| **Quick and Choo 2014** | Google Drive | Cloud Storage |
| **Farina et al. 2014** | Bittorrent | Torrent Client |
| **Blakeley et al. 2015** | hubiC | Cloud Storage |
| **Shariati et al. 2015** | SugarSync | Cloud Storage |
| **Dezfouli et al. 2015** | Facebook, Twitter, LinkdIn, Google + | Social Media |
| **Yang et al. 2016** | Facebook, Skype | Social Media, VoIP |
| **Daryabar et al. 2016** | OneDrive, Box, Google Drive, Dropbox | Cloud Storage |
| **Daryabar et al. 2016** | MEGA | Cloud Storage |

As can be seen from Table 1, there was no previous analyzing FxOS network traffic artefacts. FxOS is designed to allow mobile devices to communicate directly with HTML5 applications using JavaScript and newly introduced WebAPI. However, the used of JavaScript in HTML5 applications and solely no OS restriction might lead to security issues and further potential exploits and threats. FxOS is still not fully supported by most of the existing mobile forensic tools which further urgencies for further research development in this area (Yusoff et al., 2014a).

In this chapter, we were focused on analysis of residual network traffic artefacts in FxOS. Two popular social networking applications (Facebook and Twitter) and one instant messaging application (Telegram) were investigated as case studies. In the earliest days, investigators used to put the mobile phone in the special sandbox to monitor mobile GSM activities (Androulidakis, 2012). This method however, requires a lot of expensive devices and thus, the phone monitoring process became more complicated and very costly. FxOS simulator is a virtualised version of the

FxOS that runs provides full user experience and FxOS features. Therefore, we have followed the method proposed by Quick and Choo and simulated FxOS in a virtual environment machine (Quick and Choo, 2013c). When we performed the communication activities using the FxOS simulator, we used the Network Analyzer to capture and monitor the network traffic.

The rest of this chapter is organized as follows. In section 2, we have explained the methodology used and outlined the setup for our experiment. In section 3, we have presented our research findings and finally concluded our research in section 4.

## 2. Experiment Setup

Network analysis is a procedure for investigating the movement of data that travel across the targeted network. This procedure was performed by analyzing and carving the captured network artefacts. In general, the collection of network artefact for mobile devices is very difficult because of the limitation of mobile hardware and mobile OS itself. It is in contrary with conventional desktop OS, which we can easily captured the network files using various tools available. For that reason alone, we have adapted an approach for forensic collection of cloud artefacts (Quick and Choo, 2013b, 2013c, 2014) into our research methodology. Quick and Choo analysed several cloud applications using virtual machines and captured and analysed network traffic activities. In this chapter, we run FxOS simulator within VMware (VM) (VMware, 2013), configuerd with two popular social networking applications (Facebook and Twitter) and one instant messaging application (Telegram); which we will perform communication tasks in different scenarios. All the network activities are then captured and analysed using network analysis tools.

FxOS Simulator is an add-on simulator for the Firefox browser that enables users to run FxOS on any desktop computer. It comes with the Dashboard, a tool hosted by Firefox web-browser that enables users to start and stop the simulator; install and uninstall the applications; and to debug FxOS applications . The Dashboard is also use to push applications to a real device, and checks application manifests for any common problems. Three applications which are two popular social networking applications (Facebook and Twitter) and one instant messaging application (Telegram) installed in the VM disk and communications activities were performed to detect and investigate residential artefacts. Separated VM disks were then created for every taken action and all the network activities were captured and monitored. Figure illustrates our experiments setup.

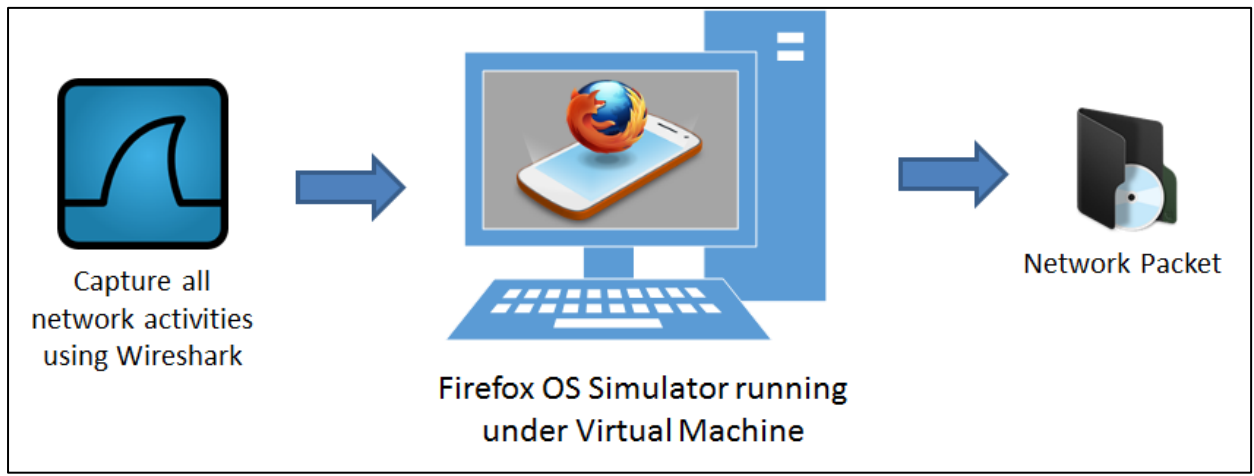

Figure 1: FxOS Network Traffic Analysis Methodology.

As the network analyzer captures all VM network activities including Windows 8.1 services we have filtered out non relevant network data using Wireshark 1.12.5. We have repeated our investigation using Ubuntu 14.04 LTS and the network packets in both environment were complementarities. This research experiment setup was divided into four stages; (1) Preparing virtual machines with selected applications; (2) Executing the activities and documenting all steps taken (3) Capturing the network activities; and (4) Conducting network analysis.

## 2.1 Preparing Virtual Machines

In this experiment, we run FxOS simulator on Windows 8.1 and Ubuntu 14.04 LTS uutilising VMware Player 10.0.1. Our experiments were mainly focused on applications that regularly stimulate user communications and applications usage. We have configured two popular social networking applications (Facebook and Twitter) and one instant messaging application (Telegram) as our test subjects. Several tasks were performed in different sets of scenarios and all the network activities were captured using Wireshark 1.12.5 and Microsoft Network Monitor 3.4. (Microsoft, 2010). For analysis part, we have used the Network Miner 1.6.1 to carve related network artefacts. Table 2 shows all the software and applications that were used in our experiment

Table 2 - Software and Applications for FxOS Network Traffic Investigation

| Software or Application | Purpose |
| --- | --- |
| Windows 8.1 | Operating System |
| Ubuntu 14.04 LTS | Operating System |
| VMware Player 10.0.1 | Provides virtual environment |
| Wireshark 1.12.5 | Capturing and monitoring network activities |
| Microsoft Network Monitor 3.4 | Monitoring network activities |
| Network Miner 1.6.1 | Carving and identifying evidences from network packets |
| Facebook | Test application (Social Media) |
| Twitter | Test application (Social Media) |
| Telegram | Test application (Instant Messaging) |

As shown in Figure 2, a total of 11 VMs 11 physical systems were created for Windows and Ubuntu respectively. Each VM disk represents a scenario in our experiment. All VM disks were configured with Windows or Ubuntu installed on 20GB virtual hard drive and equipped with 1GB RAM.

In general, the VMs were grouped into 4 groups which are Base, Facebook, Twitter and Telegram groups. The Base VM01 consists of standard OS setup. From this point, the disk was copied, Firefox browser as well as FxOS simulator were installed in VM02. The following VMs Three copies of VM02 were configured with applications of interest (Facebook (VM03), Twitter (VM06) and Telegram (VM09)) and named accordingly. The remaining VMs were created in accordance with scenario experiments out of the first VM disks of each group.

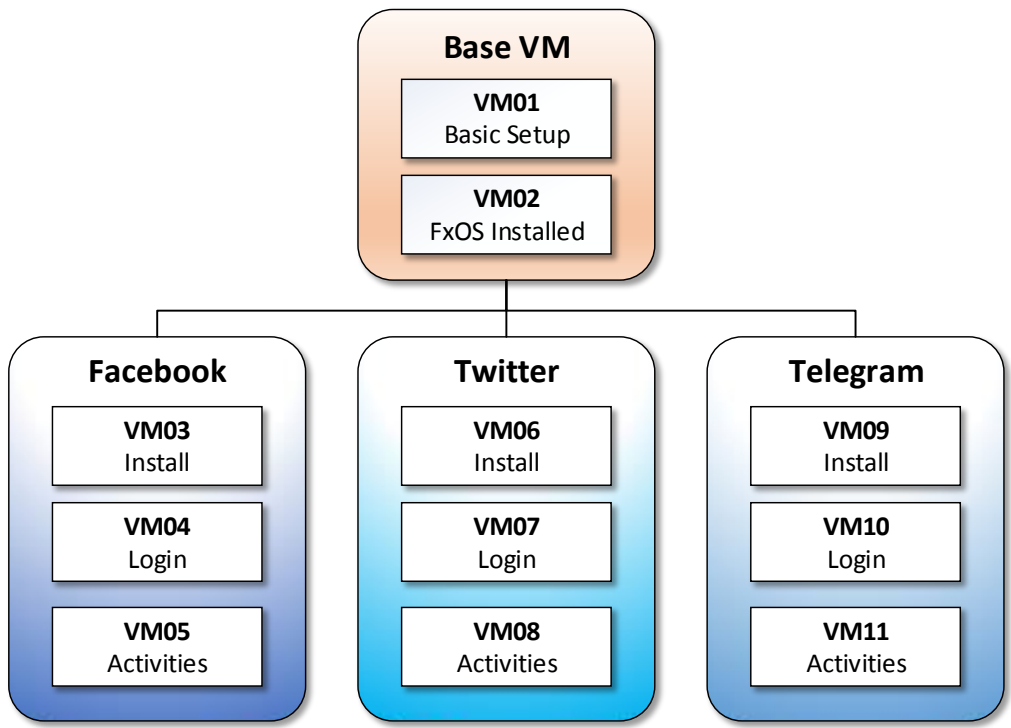

Figure 2: Virtual Machine Hierarchy for Network Analysis Investigation

## 2.2 Executing Activities

Conducting all activities and documenting all steps were taken for future reference and to support soundness of investigation. Every scenario was performed with predefined communication activities as shown in Table 3.

For login activates, we have created a dummy account with email and password of "mohd.najwadi@gmail.com" and "najwadi87" respectively. For instant messaging account, we have registered the account using mobile number "+60162444415" communication with another mobile number "+60125999159".

## 2.3 Capturing Network Activities

The objective of this analysis was to identify the correct traffic path, for several famous communication activities in mobile phone. From the packets, we need to identify what data can be seen in network traffic, what protocol were being used, who issued the certificates, and will there be any credential captured. In this experiment, we run FxOS simulator in the VMware player. All network traffics from FxOS simulator were captured once with Wireshark 1.12.5 and another time using Microsoft Network Monitor 3.4 as a backup capturing tool. In order to capture the network traffic from VMware player, VMware network adapter was set to NAT. Figure 3 shows the network packets captured by Wireshark when we executed the communication activities using the FxOS simulator.

Table 3 - Configuration of Virtual Machine with the Communication Activities

| VM Disk | Scenario | Details |
| --- | --- | --- |
| **Base** | | |
| VM01 | Base | Base configuration was installed with Windows 8.1 or Ubuntu 14.04 on 20GB virtual hard drive and 1GB Ram. |
| VM02 | Install Simulator | VM01 was copied. Firefox Browser and FxOS simulator was installed |
| **Facebook** | | |
| VM03 | Installation | VM02 was copied and Facebook application was installed |
| VM04 | Login process | VM03 was copied and used to login with prepared social media account |
| VM05 | Activities | VM04 was copied and performed posting, comment, like comment, reply comment, send message, add friend, and follow |
| **Twitter** | | |
| VM06 | Installation | VM02 was copied and Twitter application was installed |
| VM07 | Login process | VM06 was copied and used to login with prepared social media account |
| VM08 | Activities | VM07 was copied and performed tweet, reply tweet, favorite tweet, retweet, use hashtag, follow, and unfollow |
| **Telegram** | | |
| VM09 | Installation | VM02 was copied and Telegram application was installed |
| VM10 | Registration | VM09 was copied and used to register telegram account |
| VM11 | Activities | VM10 was copied and performed create contacts, send text, reply text, received picture, and share location |

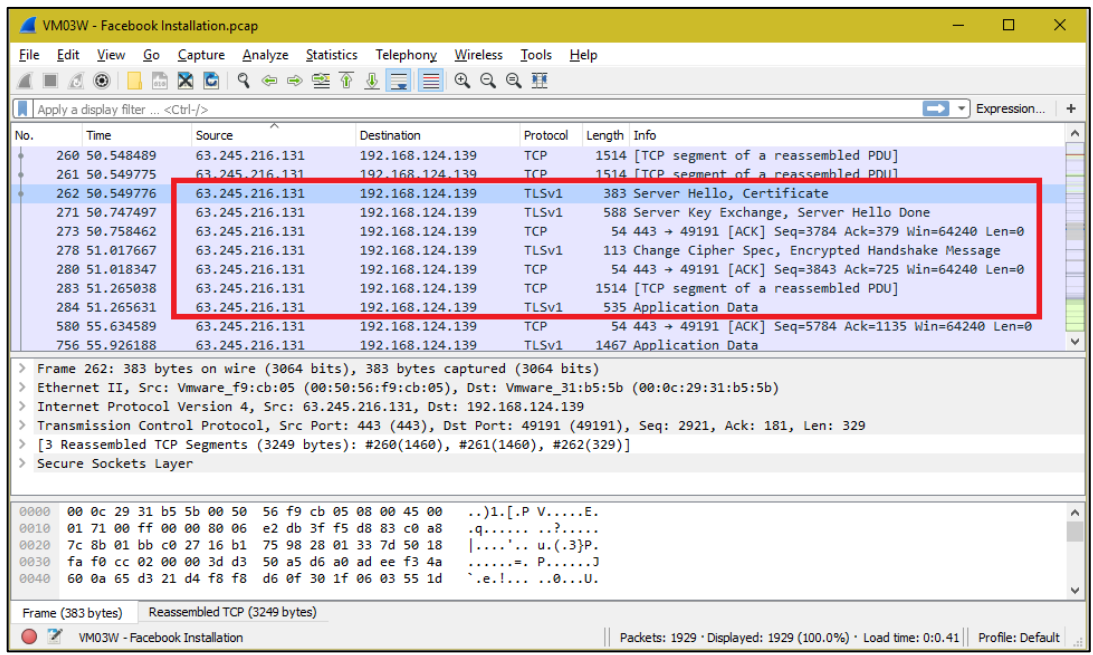

Figure 3 - Network Packets Captured by Wireshark

Apart from capturing the network activities, we have also saved the virtual memory by copying all virtual memory (.vmem) files generated by VMwares. The virtual memory files were copied after performing all the activities prior to shutting down VMs. At the end of experiments we had three files of network packets (.pcap), virtual hard drive (.vmdk) and virtual memory (.vmem) for analysis. However, this paper only reports our analysis of .pcap network traffic files and interested readers may refer to authors previous publications reporting analysis of other files (Yusoff et al., 2014a, 2014b, 2014c, 2014d).

## 2.4 Conducting Network Analysis

Network traffic analysis is the process of capturing, reviewing and analyzing network traffic for the purpose of security, performance and management. The process of analyzing the network traffic can be performed manually or using automated techniques. In this paper, network packets were analyzed manually using Network Miner 1.6.1 to find the source and destination IP address, communication port, owner of the IP, domain and subdomain, credential, images, certificate used, certificate validity, etc. Detected IP was then checked in IP address lookup website at http://www.ipchecking.com/, in the attempt to find the owner, hostname, country origin and reverse DNS.

We have also monitored and analyzed the packets using Wireshark 1.12.5 and Microsoft Network Monitor 3.4 to detect the timestamp, flow of handshakes for SSL encrypted traffics and to extract the certificate (in $\NetworkMiner_1-6-1\AssembledFiles folder according the source IP subfolder). Figure 4 shows an example of a certificate retrieved from communication with Facebook.

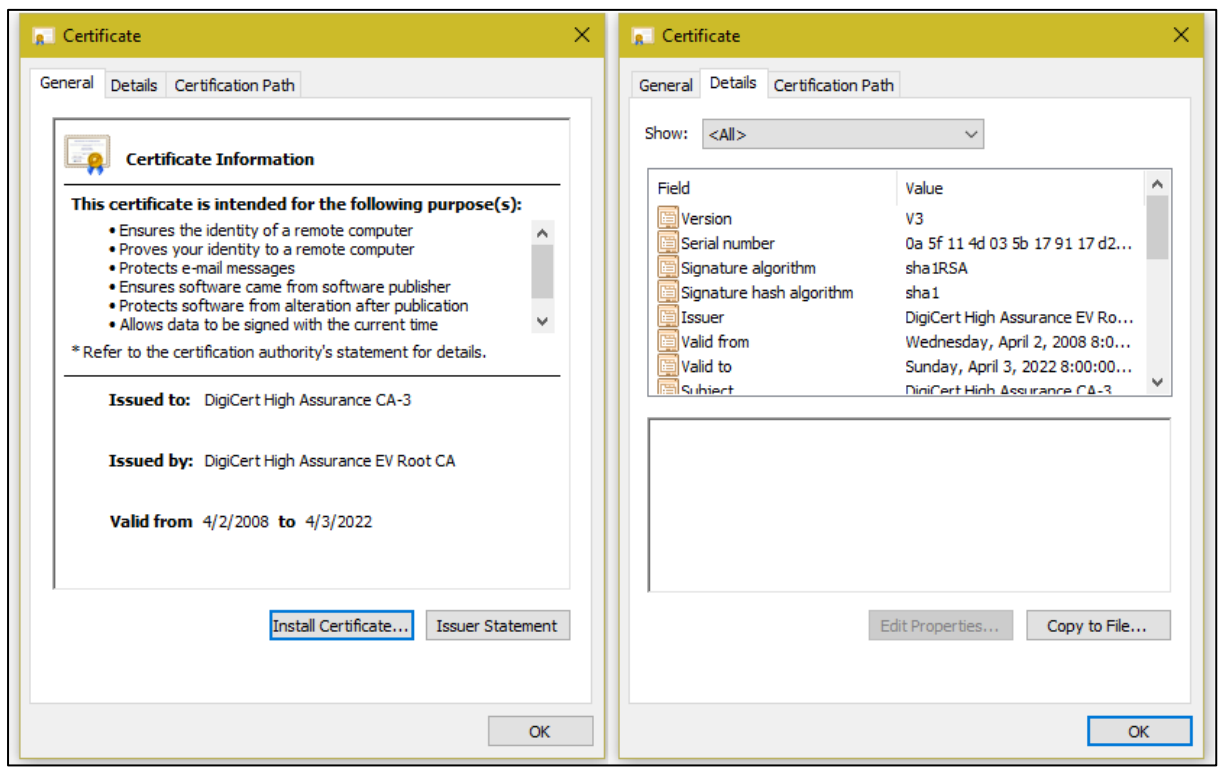

Figure 4: Captured SSL Certificate Detail Information

## 3. Discussion and Analysis

First, our network analysis started by observing the network traffic during installation of FxOS simulator. This simulator was installed as an add-on in Firefox browser. To install the simulator, we have navigated to the Firefox add-on download page, and search for FxOS simulator. The moment we click the "+ Add to Firefox" button, the initial connection was established on "download.dynect.mozilla.net" with the observed IP of 63.245.217.39 over the port 80. This IP was registered to Mozilla and we were also managed to capture all account credential with the username of "_ga=GA1.2.389580134.1432816473" without any password required for login process.

During the downloading process, we can see a high traffic movement from "*.cdn.mozilla.net" with the observed IP of 68.232.34.191 over the port 443. The connections were encrypted and the certificate was issued by DigiCert High Assurance CA-3. In addition, we have also detected network traffic from "aus4.mozilla.com" with the observed IP of 63.245.217.43, 63.245.217.138 and 63.245.217.219 over the port 80 but no data worth for forensic evidence was captured. Our next step is to observe and analyze the network movement, for each of our selected applications.

### 3.1 Network Analysis of Facebook

Facebook is one of the most used social network application in any mobile platforms. Forensic investigation of Facebook was contained very huge amount of forensic worth of evidences. Network analysis for Facebook applications consist of three stages starting from the application installation, credential login and communication activities performed using Facebook application in FxOS. VM disks were created for installation, credential login and communication activities stages for ease of organizing and monitoring purposes. When accessing FxOS marketplace, the initial session was established on "marketplace.firefox.com" with the observed IP of 63.245.216.131 over the port 80. The network movements were also detected on port 443 simultaneously and its certificate were issued by DigiCert SHA2 Extended Validation Server CA and DigiCert High Assurance EV Root CA. In addition for this case, we have also detected network traffics from Google Internet Authority with the observed IP of 216.58.210.46. These services were encrypted whereas their certificates were issued by Google Internet Authority G2, GeoTrust Global CA and Equifax Secure Certificate Authority

When browsing the application list in the marketplace, we again saw the network traffic from "*.cdn.mozilla.net" with the observed IP of 68.232.34.191 over the port 443, the same IP server when we downloaded the simulator. Once we have found Facebook application in the list, we clicked its icon and we then received the packet from "m.facebook.com" with the observed IP of 31.13.90.2 over the port 443. Twelve packets were received when we clicked the icon and the certificates were from DigiCert High Assurance CA-3 and DigiCert High Assurance EV Root CA. Next, we have captured the network traffic coming from "addons.dynect.mozilla.net" and "services.addons.mozilla.org" with the observed IP of 63.245.216.132 and 63.245.216.134 over the port 443 respectively. Both certificates were issued by DigiCert High Assurance EV CA-1. Marketplace was prompted for installation once we clicked the icon. The moment we accepted application installation, we have received the packet from "*.akamai.net" and "*.edgesuite.net" with the observed IP of 176.255.203.* over the port 80 and 443 simultaneously. This IP belongs to Facebook and the certificates were issued by Baltimore CyberTrust Root, GTE CyberTrust Global Root and Cybertrust Public SureServer SV CA. We have also detected a network traffics from "marketplace.firefox.com" and "*.cdn.mozilla.net" over the port 443 during Facebook application installation. As usual, the Facebook icon was created at the home screen as soon as the installation has finished.

Next, when we opened the Facebook application, we received the packets again from "m.facebook.com". The Facebook application has directed us to the login page and we have used the prepared Facebook account to login. The login process for Facebook application has caught our attention. Immediately after we clicked the login button, the traffic were encrypted and the packet captured were from "safebrowsing.cache.l.google.com" with the observed IP of 90.222.188.* over the port 443. The certificates were issued by Google Internet Authority G2, GeoTrust Global CA and Equifax Secure Certificate Authority. These services were used to check malicious activities which downloaded and installed malicious software without the user consent. We then repeated the login process multiple times and based on our observation, the checking were running randomly as we only manage to capture this packet twice.

After successfully authenticating to our test account, we scrolled down to the Facebook newsfeed and we have identified the captured packet again came from "*.akamai.net", "fbcdn-profile-a.akamaihd.net", and "fbcdn-photos-c-a.akamaihd.net.edgesuite.net" with the observed IP of 176.255.203.* over the port 80 and 443. Various issuer of certificates such as from Cybertrust Public SureServer SV CA, GTE Cybertrust Global Root, and Baltimore Cybertrust

Root has been identified. The packets from this server were carrying Facebook image, text, as well as other encrypted communication. We have performed several Facebook activities such as post a status, post a picture, comment and like friend's status, send Facebook private message, received private message, user search and post an emoticon. All of these activities came from "*.akamai.net", "fbcdn-profile-a.akamaihd.net", and "fbcdn-photos-c-a.akamaihd.net.edgesuite.net". On the other hand, network traffics from Google services such as Google Analytic and Google Internet Authority were captured; starting from the moment we open the marketplace until the end of our experiment. The same network traffic were also observed on Ubuntu experiment. Table 4 shows the summary of our observed IPs together with their registered organization, country origin and certificate issuers for Facebook experiment

Table 4: Observed IP and Registered Organisation for Facebook Experiment

| Registered Organization | Observed IP | Country Origin | Certificate Issuers |
|---|---|---|---|
| **Mozilla** | 63.245.216.131 | United States | - DigiCert SHA2 Extended Validation Server CA |
| | 63.245.216.132 | United States | - DigiCert High Assurance EV Root CA |
| | 63.245.216.134 | United States | |
| | 68.232.34.191 | United States | - DigiCert High Assurance CA-3 |
| **Google** | 216.58.210.46 | United States | - Google Internet Authority G2 |
| | | | - GeoTrust Global CA |
| | | | - Equifax Secure Certificate Authority |
| | 90.222.188.0 - 90.222.188.255 | United Kingdom | - Google Internet Authority G2 |
| | | | - GeoTrust Global CA |
| | | | - Equifax Secure Certificate Authority |
| **Facebook** | 31.13.90.2 | Ireland | - DigiCert High Assurance CA-3 |
| | | | - DigiCert High Assurance EV Root CA |
| | 90.223.223.0 - 90.223.223.255 | United Kingdom | - Cybertrust Public SureServer SV CA |
| | 176.255.203.0 - 176.255.203.255 | United Kingdom | - GTE Cybertrust Global Root |
| | | | - Baltimore Cybertrust Root |

## 3.2 Network Analysis on Twitter

Network analysis of Twitter application followed the same steps of Facebook application consisted of three stages starting from the application installation, credential login and communication activities using Twitter application in FxOS. VM disks were created for installation, credential login and communication activities stages for ease of organizing and monitoring purposes. We again need to access FxOS marketplace in order to install the Twitter application. The packets were captured again from "marketplace.firefox.com" with the observed IP of 63.245.216.131 over the port 80 and 443 simultaneously. For encrypted packets, the certificates were issued by DigiCert SHA2 Extended Validation Server CA and DigiCert High Assurance EV Root CA. Similarly, we have also detected network traffics from Google Internet Authority with the observed IP of 216.58.210.78 for this case. These services were encrypted whereas their certificates were issued by Google Internet Authority G2, GeoTrust Global CA and Equifax Secure Certificate Authority.

Likewise, we again saw the network traffic from "*.cdn.mozilla.net" with the observed IP of 68.232.34.191 over the port 443, when we browsed the application list in Mozilla Marketplace and the certificate was issued by DigiCert High Assurance CA-3. However, we have captured a different source of packets when we clicked on Twitter icon. The packets we received were from "mobile.twitter.com" with the observed IP of 185.45.5.37 and 185.45.5.48 over the port 443.

Both IP's certificates were issued by Symantec Class 3 Secure Server CA-G4 and Symantec Class 3 Public Primary Certification Authority-G5. We have also managed to capture the network traffic from "addons.dynect.mozilla.net" and "services.addons.mozilla.org" with the observed IP of 63.245.216.132 and 63.245.216.134 over the port 443 respectively, which were the same packets at the same stage during our previous experiment. At the point we accepted for application installation, we received the packets from "ocsp.ws.symantec.com.edgekey.net" and "ss.symcd.com" with the observed IP of 23.54.139.27 over the port 80. This server was transmitting the application data from the server hosted by Akamai Technologies, a cloud service provider based in the United States. As usual, network traffics came from Google services as well as from "marketplace.firefox.com" and "*.cdn.mozilla.net" over the port 443, were also detected during Twitter application installation. The Twitter icon was created at the home screen as soon as the installation was finished.

Next, when opening the Twitter application, we received the packets again from "mobile.twitter.com". Furthermore, we have also managed to capture the packets from "cs139.wac.edgecastcdn.net" and "*.twimg.com" with the observed IP of 68.232.35.172 over the port 443. The certificates were issued by DigiCert High Assurance EV Root CA and DigiCert SHA2 High Assurance Server CA. From our investigation, these servers belong to Twitter. Similar to Facebook application, Twitter application has also directed us to the login page and we have used the pre-prepared Twitter account to login. For Twitter application, immediately after we clicked login button, the traffic were encrypted and we also managed to capture the packet from "safebrowsing.cache.l.google.com" with the observed IP of 90.222.188.* over the port 443. The certificates were issued by Google Internet Authority G2, GeoTrust Global CA and Equifax Secure Certificate Authority. This services were used to check downloaded and installed malicious software without user consent. In contrast to Facebook, we have managed to capture this packet every time we login to Twitter application. We continually received packet from "mobile.twitter.com" along, while the login process taking place.

After successfully logged in to our test account, we then scrolled down to the Twitter newsfeed and managed to capture the packets from "*.twimg.com" with the observed IP of 199.96.57.7 over the port 80 and 443 simultaneously. Previously, we have identified the packets from "*.twimg.com" but the source IP was different. The certificates were also issued by DigiCert High Assurance EV Root CA and DigiCert SHA2 High Assurance Server CA. The packet from this server carried Twitter image, text, as well as other encrypted communication. Next, we have performed several Twitter activities such as tweet, reply tweet, retweet, follow and unfollow user, direct message, view profile, create hashtag and perform user search. All of these activities came from "mobile.twitter.com" and "*.twimg.com". Similar to Facebook application, network traffics from Google services such as Google Analytic and Google Internet Authority were captured starting from the moment we open the marketplace until the end of our experiment. Table 5 shows the summary of our observed IP together with their registered organization, country origin and certificate issuers for Twitter experiment

Table 5: Observed IP and Registered Organization for Twitter Experiment

| Registered Organization | Observed IP | Country Origin | Certificate Issuers |
| --- | --- | --- | --- |
| Mozilla | 63.245.216.131<br>63.245.216.132<br>63.245.216.134 | United States | - DigiCert SHA2 Extended Validation Server CA<br>- DigiCert High Assurance EV Root CA |
| | 68.232.34.191 | United States | - DigiCert High Assurance CA-3 |
| **Google** | 216.58.210.78 | United States | - Google Internet Authority G2<br>- GeoTrust Global CA<br>- Equifax Secure Certificate Authority |
| | 90.222.188.0 -<br>90.222.188.255 | United Kingdom | - Google Internet Authority G2<br>- GeoTrust Global CA<br>- Equifax Secure Certificate Authority |
| **Akamai Technologies** | 23.54.139.27 | United States | N/A |
| **Twitter** | 68.232.35.172<br>199.96.57.7 | United States | - DigiCert High Assurance EV Root CA<br>- DigiCert SHA2 High Assurance Server CA |
| | 185.45.5.37<br>185.45.5.48 | United States | - Symantec Class 3 Secure Server CA-G4<br>- Symantec Class 3 Public Primary Certification Authority-G5 |

### 3.3 Network Analysis on Telegram

Telegram was the only instant messaging application that we used in our experiment. following the same steps for Facebook and Twitter application, this experiment has also consists of three stages starting from the application installation, credential login and communication activities using Telegram application in FxOS. VM disks were created again for installation, credential login and communication activities stages, to ease the organizing and monitoring purposes. During installation of Telegram application, the packet were captured again from "marketplace.firefox.com" with the observed IP of 63.245.216.131 over the port 80 and 443 simultaneously when we opened Mozilla Marketplace application from FxOS simulator. For encrypted packets, the certificates were issued by DigiCert SHA2 Extended Validation Server CA and DigiCert High Assurance EV Root CA.

Next, when we browsed and searched the Telegram application in Mozilla Marketplace, we saw again the network traffic from "*.cdn.mozilla.net" with the observed IP of 68.232.34.191 over the port 443 with the certificate was issued by DigiCert High Assurance CA-3. When we clicked the Telegram icon, we have received the installation data from "download.cdn.mozilla.net" with the observed IP of 93.184.221.133 over the post 80. We have also managed to capture the network traffic from "addons.dynect.mozilla.net" and "services.addons.mozilla.org" with the observed IP of 63.245.216.132 and 63.245.216.134 over the port 443 respectively, which were the same packets at the same stage during our previous experiment. The same as previous experiment, network traffics from "marketplace.firefox.com" and "*.cdn.mozilla.net" over the port 443 have also been detected during Telegram application installation. The Telegram icon was created at the home screen as soon as the installation finished.

Next, we then opened the Telegram application and to our surprise, we have received the packet from "marketplace.firefox.com" with the observed IP of 63.245.216.131. Normally, this packet is received when we clicked on Mozilla Marketplace icon. When the application was executed, we proceeded with Telegram phone number registration. The moment the phone number and country were selected, we received a packet from "addons-blocklist-single1.vips.phx1.mozilla.com" with the observed IP of 63.245.217.113 over the port 80. Next, clicking the next button and while Telegram application was generating the registration key, we received the network traffic from "telegram.org" with the observed IP of 149.154.167.51 over the port 80. Registration code was received and our network capturing software recorded incoming network traffic again from "telegram.org" but with different observed IP which was 149.154.167.91 over the port 80. At the same time, we have also received the network traffic from "*.us-west-2.compute.amazonaws.com" with the observed IP of 52.24.145.20 over the port 443. The certificates were issued by DigiCert Global Root CA and DigiCert SHA2 Secure Server CA. When we repeated our experiment with Ubuntu, the same traffic came through when we received registration key for Telegram. Therefore, we can conclude that this server was used to push and generate the registration key for Telegram. In contrary to Facebook and Twitter network analysis, we were no longer received the network traffic from "safebrowsing.cache.l.google.com" with the observed IP of 90.222.188.* during our experiment. This is because Telegram is only involved with messaging services and no browsing mechanisms were included.

After registration process completed, we have performed several communication activities such as creating a new contact, opening chat windows, creating group and sending messages to other contact. During the activities, we have received the network traffic from "telegram.org" with the observed IP of 149.154.171.* over the port 80. When we received a message from other contact, incoming traffic were from "github.map.fastly.net" with the observed IP of 185.31.19.133 over the port 443. The certificates were issued by DigiCert High Assurance EV Root CA and DigiCert SHA2 High Assurance Server CA. We then continue our experiment by sharing our location to our contact and the network traffic was recorded from Google Internet Authority with the observed IP of 216.58.209.234 over the port 443. The services were encrypted whereas their certificates were issued by Google Internet Authority G2, GeoTrust Global CA and Equifax Secure Certificate Authority. Finally we ended our experiment by playing the song received from other contact and again, we received the network traffic from "*.us-west-2.compute.amazonaws.com" with the observed IP of 52.24.145.203 over the port 443. Table 6 shows the summary of our observed IP together with their registered organization, country of origin and certificate issuers for Telegram experiment.

Table 6: Observed IP and Registered Organisation for Twitter Experiment

| Registered Organization | Observed IP | Country Origin | Certificate Issuers |
|---|---|---|---|
| **Mozilla** | 63.245.217.113 | United States | N/A |
| | 63.245.216.131<br>63.245.216.132<br>63.245.216.134 | United States | - DigiCert SHA2 Extended Validation Server CA<br>- DigiCert High Assurance EV Root CA |
| | 68.232.34.191 | United States | - DigiCert High Assurance CA-3 |
| | 93.184.221.133 | United States | N/A |
| **Google** | 216.58.209.234 | United States | - Google Internet Authority G2<br>- GeoTrust Global CA<br>- Equifax Secure Certificate Authority |
| **Telegram** | 149.154.167.51<br>149.154.167.91 | United States | N/A |
| | 149.154.171.0 - 149.154.171.255 | United States | N/A |
| **Github** | 185.31.19.133 | Unknown | - DigiCert High Assurance EV Root CA<br>- DigiCert SHA2 High Assurance Server CA |
| **Amazon** | 52.24.145.20<br>52.24.145.203 | United States | - DigiCert Global Root CA<br>- DigiCert SHA2 Secure Server CA |

# 4. Conclusion and Future Works

Network analysis is a vital piece in conducting mobile forensics. In this paper, we have successfully presented the network analysis of two popular social networking applications (Facebook and Twitter) and one instant messaging application (Telegram) on FxOS. The findings of this study reported many valuable forensics evidences such as image files, communication texts and authentication credentials detectable in the network traffic. Fortunately, captured credentials were not in plaintext. Communications in Telegram were transmitted over port 80 in plain text. All communication activities in Facebook and Twitter however were encrypted and transmitted over port 443. The other conclusion drawn from this research was that not all service providers are storing client data on their servers i.e. Twitter is using the cloud services from Akamai Technologies to store their installation files. Multiple certificates were carved from the packets namely Mozilla used the certificates from DigiCert; Google from Google Internet Authority, Facebook and Twitter from DigiCert.

This research has thrown up many questions in need of further investigation. More information of FxOS applications investigation would help us to establish a greater degree of network traffic forensic accuracy. Further research opportunities include undertaking the process outline in this research for cloud storage services. Previous forensic investigation on cloud storage services generally used the cloud storage applications on Apple iOS and Google Android as case studies. Therefore, the presence of FxOS will increase in-depth study of FxOS forensic in cloud storage forensic area. In addition, this research is the first forensic investigation to use the phone simulator in order to monitor network traffic on mobile phone. A future study of investigating network traffic on other mobile OS using phone simulator would be very interesting.